# Phase field model of Coulomb explosion damage in solid induced by ultrashort laser

A○, L̥ɪʜᴏɴɢ,[1,†,*] ᴀɴᴅ Zʜᴀɴɢ, Qɪɴ[1,†]

[1]*College of Physics and Optoelectronic, Shenzhen University Shenzhen 518060, P. R. China*
[†]*These authors contributed equally to this work.*
* aolihong2019@email.szu.edu.cn

Much experimental evidence reveals that Coulomb explosion governs non-thermal material removal under femtosecond or even shorter laser pulses, and non-thermal laser damage has been a topic widely discussed. Nevertheless, there is still no continuum mechanical model capable of describing the evolution of such damage. In this study, we develop a model that characterizes solid damage through a phase field variable governed by Allen-Cahn dynamics. The parameter of the model is defined by a conceptual mechanism: during Coulomb explosion, electron pressure surpasses the interatomic barrier potential, dissociates material from the solid surface as small equivalent particles and resulting in localized damage. The numerical simulation validates the model's availability and demonstrate its ability to predict damage morphology under varying laser conditions. This work advances the understanding of non-thermal ablation and provides a tool for optimizing ultrafast laser processing.

## 1.  Introduction

The interaction between ultrashort laser pulses and solid materials has been extensively studied across various scientific and industrial domains, such as material processing [1]–[4], medical applications [5]–[7] and energy sector [8] since the beginning of the century. This interaction primarily involves the removal of material from a solid surface through laser irradiation, resulting in rapid heating and diverse phase transitions. A comprehensive understanding of the underlying physical mechanisms of laser ablation and the development of accurate mechanical models are crucial for optimizing the efficiency and precision of laser-based applications.

The description paradigm of the ultrashort laser-solid interactions primarily falls into two categories: molecular dynamics and continuum dynamics. Molecular dynamics describes the motion of every atoms [9]–[11], enabling the resolution of phenomena beyond the continuum limit [12], [13], but computationally demanding for practical application. By contrast, continuum models for ultrashort laser interactions combine macroscopic theories like thermo-fluid dynamics

[6], [14]–[16] and plasma physics [17]–[19]. These models describe all physical processes using partial differential equations and their boundary conditions, which are particularly advantageous for simulations at or above the micron scale. Some continuum models attribute ablation to melting and vaporization induced by temperature rise [20]–[25]. However, experimental evidence suggests that the laser input energy can be insufficient to melt or vaporize the material within the ablation crater, especially at lower laser energy densities [23], [26]–[29]. In such cases, non-thermal solid-solid phase transition is observed [30]–[34], and determining material damage based on excited electron density is more appropriate [35], [36]. High excited electron density results in Coulomb explosion [37]–[40]. In this context, a hydrodynamic Coulomb explosion model has been formulated [41]. However, no continuum model currently exists to describe Coulomb explosion from the perspective of the non-thermal solid-solid phase transitions.

In this study, we develop an electrodynamics-phase field model to characterize the damage induced by Coulomb explosion. The model is formulated based on nonlinear Maxwell equations, electron rate equation, and the Allen-Cahn equation. The phase field, governed by either the Allen-Cahn or Cahn-Hilliard dynamics, is capable of describing not only conventional phase transitions—such as those between solid, liquid, and gas [42]–[44], but also generalized phase transitions, including crystal growth [45]–[48], ferroelectric polarization [49] and structural damage [50]–[53]. The core of constructing the damage phase field (DPF) model lies in the formulation of the free energy functional. In our approach, the electron internal energy, derived from Sommerfeld's theory, establishes a relationship between the phase field free energy and the electron density and temperature. Additionally, we propose that the solid damage is equivalent to small particles of the same size and shape leaving the solid, with the free energy release attributed to the internal energy of excited electrons gas. This approach addresses the limitation of insufficient laser energy for melting or vaporizing the solid and yields a physically consistent asymmetric double-well bulk free energy function. We validated our model through a numerical example involving femtosecond laser irradiation on the surface of a silica crystal.

## 2. Model and numerical verification
### 2.1 Overview

In our model, the evolution of the DPF depends on key physical quantities, including electron density, electron temperature, and lattice temperature. Figure 1 illustrates the physical processes and their coupling relationships as described by the model. The physical processes induced by ultrashort laser irradiation of solids can be categorized into electrodynamic processes and temperature-dependent processes based on their respective time scales [54]. Electrodynamic processes, which occur within $10^{-11}$ s after a $10^{-13}$ s laser pulse, encompass the propagation of electromagnetic fields in the medium, ionization and heating of electrons, and electron-phonon coupling [55]. These processes are described by the nonlinear Maxwell equations, the electron rate equation, and the two-temperature model, respectively [35], [54], [56]–[58]. Temperature-dependent processes, on the other hand, include continuum heat transfer [56], mechanical wave propagation [58], and thermal hydromechanical behaviors such as fluid flow [59], melting, and vaporization. Unlike conventional thermal hydromechanical phase transitions (e.g., melting and vaporization), the evolution of the DPF in Coulomb explosions operates on an electrodynamic time scale[60], [61]. This is because the variations in electron density and electron temperature, which drive Coulomb explosions, typically occur over extremely short time scale. It is noteworthy that the model uses one-way coupling to describe electromagnetic and temperature fields influencing the DPF, which reducing complexity but limiting its use for long or multi-pulse lasers where two-way coupling effects matter.

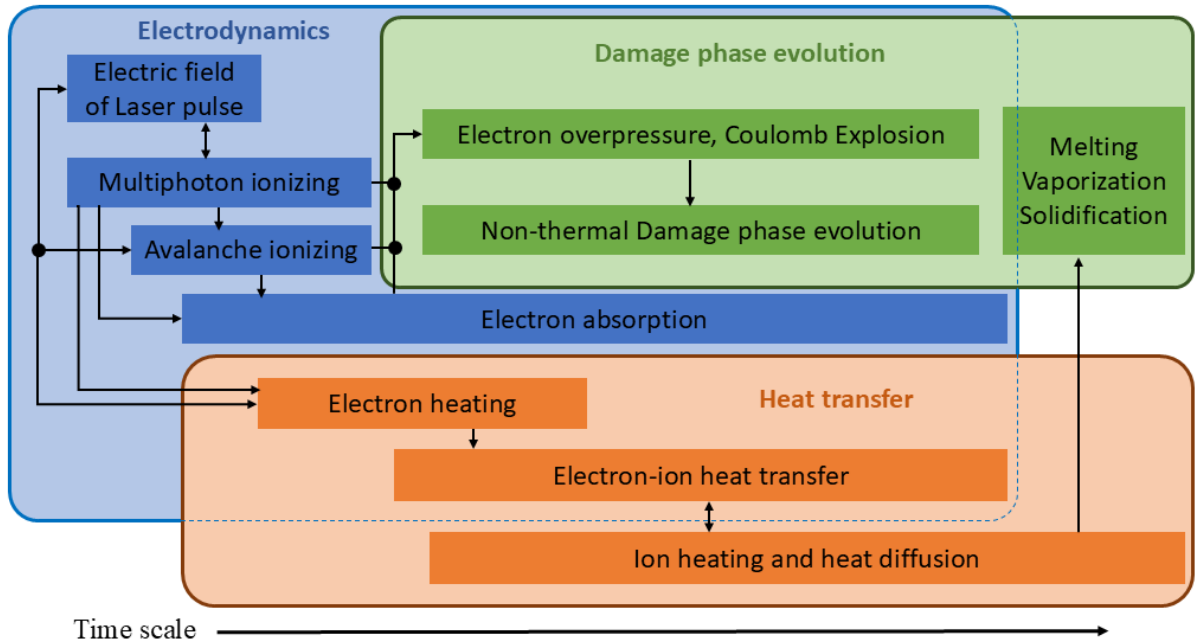

*Figure 1 physical processes described by our model. Time evolves from left to right. Explicit relations between physical processes are indicated by arrows. Black circular nodes indicate interconnections between arrows.*

### 2.2 Ionization-heating model

Before introducing the DPF model, we first present the Ionization-heating (I-H) model and two-temperature model, and derive the physical quantities required by the DPF model. The used parameter values in our numerical example are directly indicated in the text. First, the electromagnetic field $\boldsymbol{E}, \boldsymbol{H}$, polarization current density $\boldsymbol{J}_e$ and other current density, electron density $n_e$ are described by the ionization model comprises the nonlinear Maxwell equation and the rate equation for free electrons:

$$\begin{cases} \dfrac{\partial \boldsymbol{E}}{\partial t} = \dfrac{1}{\epsilon_0 \epsilon_r}\left(\boldsymbol{\nabla} \times \boldsymbol{H} - \boldsymbol{J}_e - \epsilon_0 \chi_3 \dfrac{\partial(|\boldsymbol{E}|^2 \boldsymbol{E})}{\partial t} - \dfrac{E_g w_{pi} \boldsymbol{E}}{I}\dfrac{n_a - n_e}{n_a}\right) \\ \dfrac{\partial \boldsymbol{H}}{\partial t} = -\dfrac{1}{\mu_0}\boldsymbol{\nabla} \times \boldsymbol{E} \\ \dfrac{\partial \boldsymbol{J}_e}{\partial t} = -\dfrac{\boldsymbol{J}_e}{\tau_e} + \dfrac{e^2 n_e}{m_e}\boldsymbol{E} \\ \dfrac{\partial n_e}{\partial t} = \left(w_{pi} + \dfrac{4\log 2\, e^2 \tau_e |\boldsymbol{E}|^2}{3 m_e E_g (1+\omega^2 \tau_e^2)} n_e\right)\dfrac{n_a - n_e}{n_a} - \dfrac{n_e}{\tau_{rec}} + D_e \nabla^2 n_e \end{cases} \quad (1)$$

Where $I = \dfrac{1}{2}\sqrt{\dfrac{\epsilon_0 \epsilon_\infty}{\mu_0}}|\boldsymbol{E}|^2$ is wave intensity. For an incident wave with $\lambda = 600$ nm wavelength, $\omega = 2\pi\dfrac{c}{\lambda}$ is the angle frequency. For silica, $\tau_e = 0.5$ fs [55] and $\tau_{rec} = 6$ ps are electron collision period and electron recombination period in Drude model. $m_e, e$ are mass and charge of an electron. $E_g = 9$ eV is electron band gap. $\epsilon_r = 2.105$ takes from the non-excited fused silica. $n_a = 2 \times 10^{28}$ m$^{-3}$ is the saturation density. The third term of $\partial \boldsymbol{E}/\partial t$ is current induced by Kerr effect, in which $\chi_3 = 2 \times 10^{-22}$ m$^2$V$^{-2}$ [62]. The fourth term of $\partial \boldsymbol{E}/\partial t$ is photoionization current, in which $w_{pi} = 2 \times 10^{25}$(cm$^{-3}$s$^{-1}$) $I^6$ is photoionization rate in multiphoton range[35]. The second term of $\dfrac{\partial n_e}{\partial t}$ simulates avalanche ionization, which is a major contributor to electron density. Diffusivity $D_e = 5400/T$ cm$^2$s$^{-1}$ [63][64], $T$ is solved by the following equation (2).

Heat generation and transfer of electron and lattice is described by two-temperature model of the following form:

$$\begin{cases} \dfrac{3}{2}k_B n_e \dfrac{\partial T_e}{\partial t} = \boldsymbol{\nabla} \cdot \left(\dfrac{2 k_B^2 \mu_e n_e T_e}{e}\boldsymbol{\nabla} T_e\right) - \dfrac{3}{2}\dfrac{k_B n_e}{\tau_{ei}}(T_e - T) + \dfrac{4\pi I}{\lambda}\sqrt{\left(\sqrt{\epsilon_1^2 + \epsilon_2^2} - \epsilon_1\right)/2} \\ \rho C_i \dfrac{\partial T}{\partial t} = \boldsymbol{\nabla} \cdot (k_i \boldsymbol{\nabla} T) + \dfrac{3}{2}\dfrac{k_B n_e}{\tau_{ei}}(T_e - T) \end{cases} \quad (2)$$

where

$$\epsilon_1 = \epsilon_r - \dfrac{e^2 n_e}{m_e\left(\omega^2 + \dfrac{1}{\tau_e^2}\right)}, \quad \epsilon_2 = \dfrac{e^2 n_e}{m_e \omega \tau_e\left(\omega^2 + \dfrac{1}{\tau_e^2}\right)}$$

are respectively the real part and imaginary part of dielectric permittivity derived from Drude model. $\rho = 2203$ kg/m$^3$ and $C_i = 703$ J/kg/K is the density and heat capacity of silica. Electron-lattice relaxation time $\tau_{ei} = 1$ ps [58]. Electron mobility $\mu_e = 3 \times 10^{-5}$ m$^2$/V/s [65].

The compute domain in the following simulation is a two-dimensional rectangle, the horizontal and vertical coordinates are $x, y$. For equation (1), scattering boundary condition is used to describe $\boldsymbol{E}$ and $\boldsymbol{H}$ on the top and bottom boundary[66]–[68]. $\boldsymbol{n} \times \boldsymbol{E} = 0$ is set to the other two boundaries, $\boldsymbol{n}$ is the boundary basis. A Gaussian-type laser comes in from above, on the top boundary we have electric field in $x$ direction with intensity

$$E_x = \sqrt{\frac{2I_0}{cn_0\epsilon_0}} e^{-\frac{2x^2}{w_0^2} - \frac{(t-t_p)^2}{\Delta t^2}} \sin(\omega t) \tag{3}$$

$I_0 = 20$ TW/cm$^2$, $\Delta t = 150$ fs, $w_0 = 2$ μm, $t_p = 800$ fs are the peak intensity, time correspond to peak intensity, spatial beam radius, and temporal beam duration, respectively. $n_0 = 1.45$ takes from the non-excited fused silica. For $\boldsymbol{J}_e$ and $n_e$, no-flux condition is set on all boundaries. For $T_e$ and $T$ in equation (2), adiabatic condition is set on all boundaries. At $t = 0$, there is no electromagnetic field or current in the solution domain, $n_e = 10^{-6}$ nm$^{-3}$, and $T_e = T = 300$K.

### 2.3 Damage phase field model

The physical processes illustrated in Figure 2 can be summarized as follows. The laser pulse ionizes the solid, generating a region of high free electron density. The resulting electron pressure within the ionized solid drives the Coulomb explosion, which causes emission of solid components and disorder of the lattice, finally leads to the expansion of the damage region. Solid heating occurs subsequently. The solid relaxation to a more stable state and cause damage is conceptualized as a micro-particle emission process. The particles termed "equivalent particles", this is because Coulomb explosion experiments observe not only nano-particle dissociated from a solid surface [69], electron and ion emissions [37], [70], [71] but also crystallization mechanisms [31]. When the equivalent particle is separated from the solid, it will be able to release the electron pressure through ionization, reducing the free energy. Electron pressure gradient $\boldsymbol{\nabla} p_e$ drives equivalent particle dissociation, propagating $p_e$ migration along $\boldsymbol{\nabla} p_e$, and continuing dissociation. This process has many similarities to the hydrodynamic model [41] in a mechanical perspective. But unlike the hydrodynamic model, which are concerned with the motion of the emitting medium, our model is concerned with the morphology of the damaged solid interface.

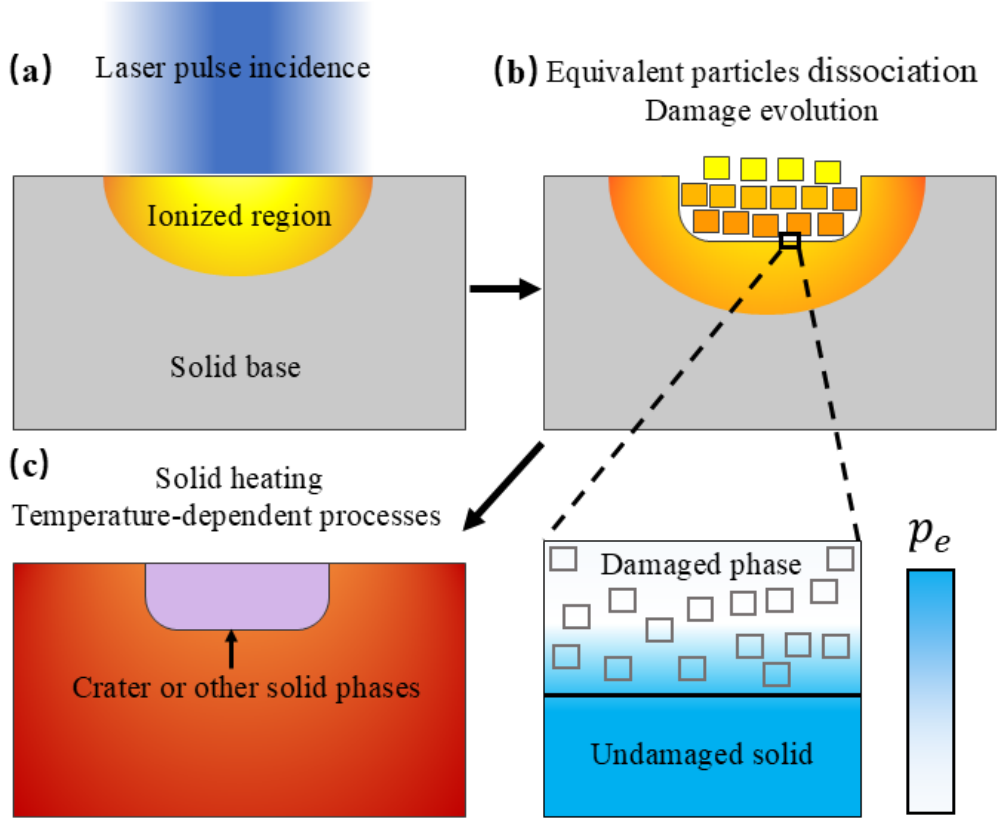

*Figure 2 Schematic drawing of the mechanical processes described by the ionization-heating model and damage phase field model. (a) Laser incidence and ionization. (b) Equivalent particles dissociate from solid. The subgraph indicates the distribution of the effective electron pressure gradient at the damage interface. (c) Damage phase produced, the heat transfer continues.*

Our DPF model specifically focuses on the Allen-Cahn dynamics governing the damage evolution process. The free energy functional $F$ on domain $\Omega$ is depending on a phase field $\phi$ and other fields $n_e, T_e, T$ solved by the I-H model and two-temperature model above:

$$F = \int_\Omega (f(\phi, n_e, T_e, T) + \kappa|\nabla\phi|^2)\, dV \tag{4}$$

$\kappa$ is an interface thickness-controlled coefficient. Like solidification[72], we suppose the specific form of free energy density $f(\phi, n_e, T_e, T)$ as an asymmetric double well function of $\phi$:

$$f(\phi, n_e, T_e, T) = \left(\alpha(\phi-1)^2 + \beta\left(\phi - \frac{3}{2}\right)\right)\phi^2 \tag{5}$$

Substituting into the general form of Allen-Cahn equation $\frac{\partial\phi}{\partial t} = -M\frac{\delta F}{\delta \phi}$ yields

$$\frac{\partial\phi}{\partial t} = -M\big(\phi(\phi-1)(3\beta + \alpha(4\phi - 2)) - \kappa\nabla^2\phi\big) + S_{nuc} \tag{6}$$

Where $\alpha = \alpha(n_e, T_e, T)$ and $\beta = \beta(n_e, T_e, T)$ reflects the barrier height and the energy released by the phase transition. $M$ is the mobility coefficient. Gradient coefficient $\kappa = 1.36\,\sigma l_i$ [73], where $\sigma$ is the interfacial energy density, and $l_i$ is the phase interface thickness. To satisfy the continuum approximation, we take $l_i = 10a$, lattice constant $a = 0.5$nm. Source term $S_{nuc}$ simulates phase nucleation.

This section builds a simple model to deduce the parameters including $\alpha$, $\beta$ and $M$ in equation (6). The following assumptions & approximations are made:

(1) The internal energy $\epsilon_e$ and pressure $p_e$ of the excited electron can be described by Sommerfeld's electron gas theory.
(2) The damage phase is consisted of equivalent particles leaving the solid, which are equivalently considered to be of the same shape and size.
(3) The equivalent particles release all the excited electronic pressure when removed from the solid.
(4) There exists a *critical distance* $\chi$, when the equivalent particles are at this distance from the solid surface, the electron repulsion pressure between them jumps from $p_e$ to 0.

The electrons excited above the conduction band comprise an electron gas that satisfies assumptions & approximations (1), which has internal energy density $\epsilon_e = \frac{3}{2}\epsilon_F n_e \sqrt{(\frac{k_B T_e}{\epsilon_F})^2 + (\frac{2}{5})^2}$ and pressure $p_e = \frac{2}{3}\epsilon_e$ [41], [74], [75]. $\epsilon_F = \frac{\hbar^2}{2m_e}(3\pi^2 n_e)^{2/3}$ is the Fermi energy of electron gas. $\epsilon_e$ can be reduced to

$$\epsilon_e = \frac{3n_e}{2}\sqrt{\left(\frac{3^{2/3}\pi^{4/3}\hbar^2 n_e^{2/3}}{5m_e}\right)^2 + (k_B T_e)^2} \tag{7}$$

Although the DPF model cannot predict the microscopic morphology of the damaged phase, the specific microscopic morphology of damaged phase will determine the value of phase field parameter values. Defining the energy ratio of the chemical bonds to be broken for the particle to leave the solid as $\eta$. Considering the particles of the cube with length of the side $l$, which is several times or more larger than the lattice constant. Each unit cell contains 12 Si-O bonds, with surface cells having 2 broken bonds. Thus, the energy ratio can be written as $\eta \approx 6l^2\frac{2}{a^2}/(12\,l^3/a^3) = a/l$. While the energy that breaks all the bonds of the crystal per unit volume can be regarded as the heat of vaporization, which is approximately equal to

$$\epsilon_s = \int_T^{T_{boiling}} C_i \rho\, dT \approx 1.54 \times 10^6(3220 - T) \tag{8}$$

The unit of $\epsilon_s$ and $T$ is J/m³ and K. The barrier energy density is $\eta\epsilon_s = al\epsilon_s$. According to assumptions & approximations (3), free energy density decreases $\epsilon_e - \epsilon_s$. Then, for the maximum of $f(\phi)$ ($\phi \in [0,1]$) to be $al\epsilon_s$, and $f(1) = -(\epsilon_e - \epsilon_s)$, then $\alpha, \beta$ must satisfy

$$\alpha = 16al\epsilon_s$$
$$\beta = 2(\epsilon_e - \epsilon_s) \tag{9}$$

Because the electron density and electron temperature the phase field depends on changes rapidly with time, the value of M will affect the final phase field evolution result. The value or specific form of $M$ is set as a constant[32]–[34], [56], a function of the curvature of phase interface[76], or other configurations[72], [77] when building a DPF model. It is the ratio of interface migration speed $v_i$ to interface energy $\sigma$. With fracture energy 0.81 GPa [78] and the length of Si-O bond 0.16 nm, $\sigma$ of silica is on the order of 0.1 J/m², we take 0.13 J/m². The critical distance $\chi$ in assumptions & approximations (4) reflects the limits of the interaction distance between particles and solids. Energy is released to accelerate the particles to speed $v_0$ within this distance. That is to say $\rho l \frac{v_0^2}{2} = p_e \chi$ and $v_i = v_0/\chi$. Sort out the above relations, we have

$$M = \frac{1}{\sigma}\sqrt{\frac{2n_e l}{\rho \chi}}\left(\left(\frac{3^{2/3}\pi^{4/3}\hbar^2 n_e^{2/3}}{5m_e}\right)^2 + (k_B T_e)^2\right)^{1/4} \tag{10}$$

The nucleation $S_{nuc}$ inside the solid is not considered in our model, that is, nonzero nucleation source is only applied on the initial solid surface. $S_{nuc}$ will be reflected in the boundary conditions of the equation (6). We set the nucleation condition to be $\beta > 0$, this is when the release of the free energy of the electron gas is greater than the barrier to be overcome for the formation of the damaged phase. $S_{nuc}$ in the following simulation is regard as a boundary flux

$$-\boldsymbol{n} \cdot (\boldsymbol{\nabla} \cdot S_{nuc}) = \begin{cases} 10^{-6} \text{ m/s/Pa} \cdot (1-\phi)\beta, & (\beta > 0) \\ 0, & (\beta \leq 0) \end{cases} \tag{11}$$

Its effect is to quickly raise $\phi$ from 0 to 1 when $\beta > 0$ is locally satisfied on the solid surface.

With initial condition $\phi = 0$, Allen-Cahn equations (6) with formula (7) ~ (10) substituted and the boundary condition (11) constitute the final form of the problem that our model is trying to solve.

### 2.4 Numerical implement

The above equations are solved in a 5 μm × 5 μm two-dimensional computation domain using the finite element method implemented in COMSOL Multiphysics® V6.2 software. The coupling relationships and solving order of the fields are guided by the framework illustrated in Figure 1. Since the influence of phase transitions on the I-H fields is neglected, the bidirectional coupling between these fields is eliminated. Consequently, the I-H model and the DPF model can be solved sequentially rather than simultaneously. For the I-H model, a structured square mesh with a scale of $\lambda/n_0/8$ is applied. The shape functions for the fields in Equation (1) are quadratic, while those for $T_e, T$ are linear. A segregated solver is required to ensure convergence when solving the I-H model. The DPF is discretized using an unstructured mesh with mesh size $2l_i$ and Lagrange quadratic shape functions.

## 3. Results & Discussion
## 3.1 Results of the I-H model

Figure 3 illustrates the quantities on which the evolution of the DPF depends, as obtained from the I-H model. The input settings include a peak intensity occurring at $t = t_p = 800$ fs. Shortly thereafter, the electron density $n_e$ rapidly reaches its maximum value at $t = 860$ fs, driven by strong avalanche ionization and multi-photon ionization. Figures 3(a) and 3(b) depict the spatial distributions of $|E|$ and $n_e$ at $t = 860$ fs. The nonlinear Maxwell equations predict a self-defocusing effect of the laser [79], causing the region of highest intensity to shift from the center toward the periphery. This results in the ionization zone spreading laterally along the surface rather than penetrating deeper into the material. The gradient distribution of $n_e$ along the y-direction is relatively uniform, as evidenced by the significant displacement of three contours with similar values in Figure 3(b). Consequently, using $n_e$ alone to characterize the damage induced by Coulomb explosion proves insufficient due to its inherent ambiguity. Figure 3(c) shows the lattice temperature $T$, which rises from 300K at $t = 10$ ps. Similar to the distribution of $n_e$, the maximum value of $T$ also occurs on the surface along the central symmetric line, yet it remains below the melting point of silica. This indicates that solid damage cannot be adequately explained by melting or vaporization at this stage. The temporal evolution of the spatial maxima of $|E|, n_e, T_e$ and $T$ is presented in Figure 3(d). These four quantities reach their peak values sequentially. The heating time of electrons is extremely short, with $T_e$ peaking at $t = 950$ fs, only 90 fs after the peak of $n_e$. In contrast, the time scale for electron-ion heat transfer is much longer, on the order of 1ps.

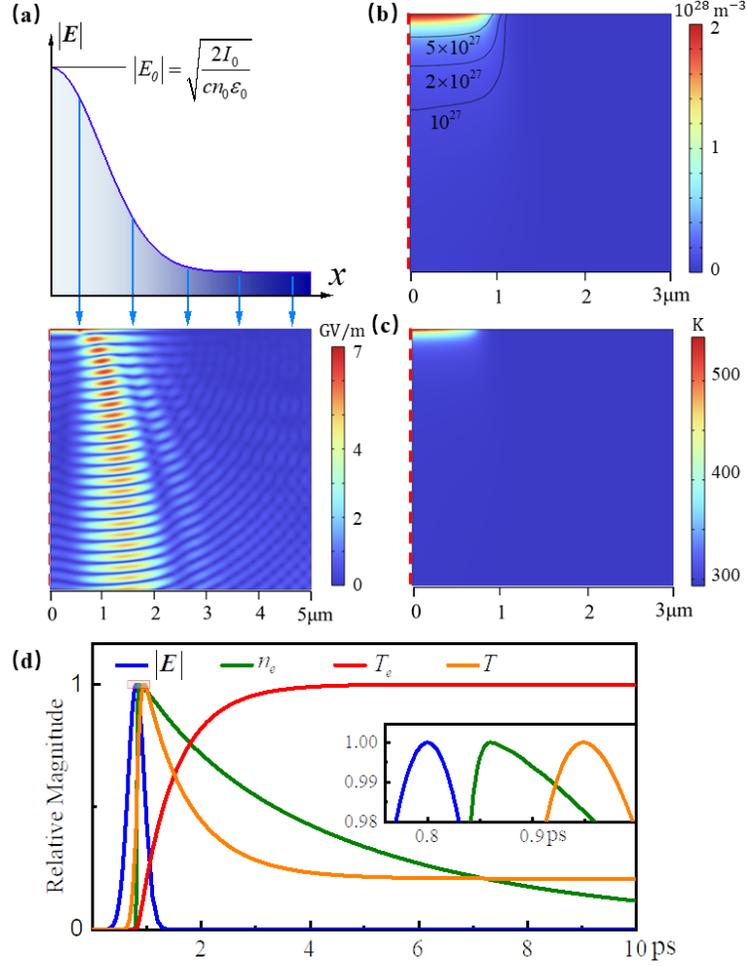

*Figure 3 (a) contour map of electric field magnitude $|E|$ at $t = 860\ fs$. Laser intensity distribution is shown above the contour map. (b) Electron density $n_e$ at $t = 860\ fs$ and (c) lattice temperature $T$ at $t = 10\ ps$. The red dashed line on the left side of the contour map is the central symmetric line. (d)Spatial maximum $|E|, n_e, T_e$ and $T$ evolve in time, the subplot is the magnified view of $|E|, n_e, T_e$ near their peak.*

## 3.2 Process of damage phase evolution

The damage phase evolution is sensitive to the free energy density drop $\beta(n_e, T_e, T)$ of the phase transition. Note that $\beta$ is independent of the phase field $\phi$, it can be derived directly from the fields solved by the I-H model. As shown in Figure 4(a)-(c), as the electron density and electron temperature increase, $\beta$ on the solid surface rises from the order of $-0.1\ J/mm^3$ to $1\ J/mm^3$, followed by a rapid decline as electron density and temperature decrease. The evolution of the damage phase also depends on two key parameters: the equivalent emitted particle size $l$ and the critical distance $\chi$. Figure 4(d)-(f) illustrates the evolution of $\phi$ when $l = 30\ nm$ and $\chi = 0.1\ a$. When the condition $\beta > 0$ is satisfied, $\phi$ begins to expand rapidly, forming a smooth phase interface, and grows to nearly half of its final distribution area in less than 100 fs. the phase interface migration rate slows significantly, almost ceasing, and the interface begins to roughen.

Figure 4 (g) shows the temporal evolution of the damage phase area $A = \int_\Omega \phi \, dS$ for different values of $l$ and $\chi$, while Figure 4(h) displays the corresponding $dA/dt$ values. Since $\beta$ is independent of the Allen-Cahn equation parameters $l$ and $\chi$ all phase fields begin to grow once $\beta > 0$. A larger $\chi$ results in a longer time for the DPF to reach a stable state. Furthermore, the evolution speed is approximately proportional to $\chi^{1/2}$, as suggested by Equation (10). A larger $l$ also accelerates DPF evolution, although its effect is less pronounced. Although Equation (6) does not guarantee that $A(t)$ is monotonically increasing, Figures 4(g) and 4(h) demonstrate that the DPF tends to stabilize over time, almost no signs of shrinkage is observed. This is because the interface mobility $M$ is decreases with $n_e$ and $T_e$, reaching a small value after several picoseconds. Consequently, the time scale of DPF evolution is on the order of electron-ion heat transfer or an order of magnitude longer than the laser pulse, approximately 1 ps here, and slightly longer than the decay time of $T_e$ under the current parameters. This agrees with experimental [32], [61] and other theoretical calculation [39], [41], [60] results on Coulomb explosions and laser induced non-thermal solid-solid phase transition. Notably, this evolution is significantly faster than the propagation of mechanical waves, as well as the phase transitions such as melting and vaporization.

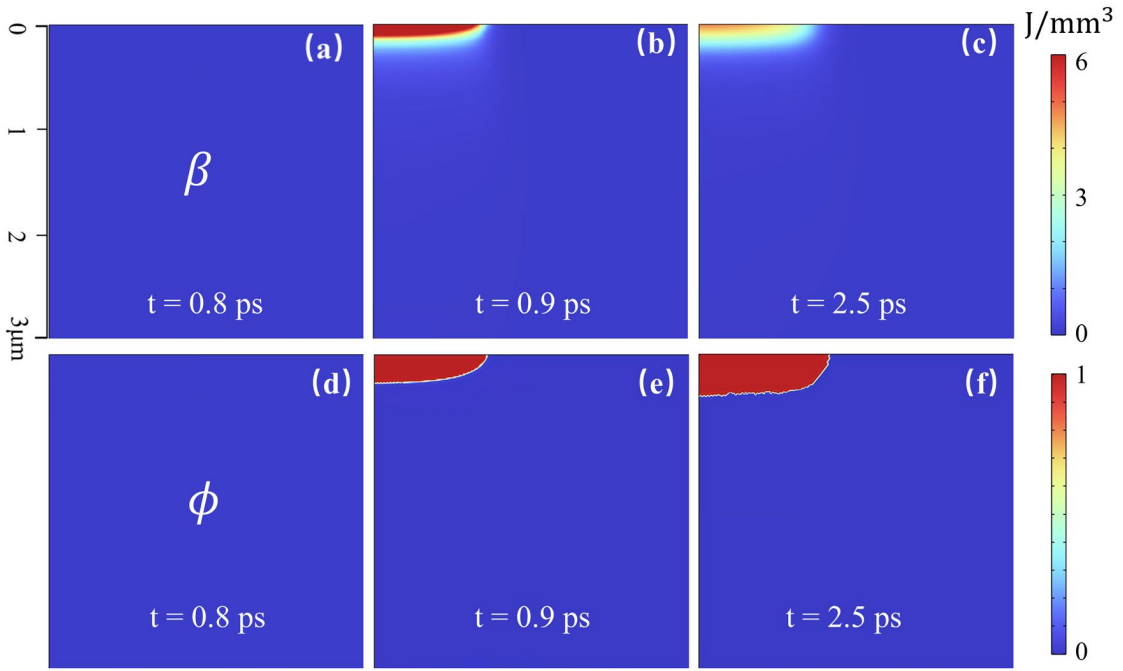

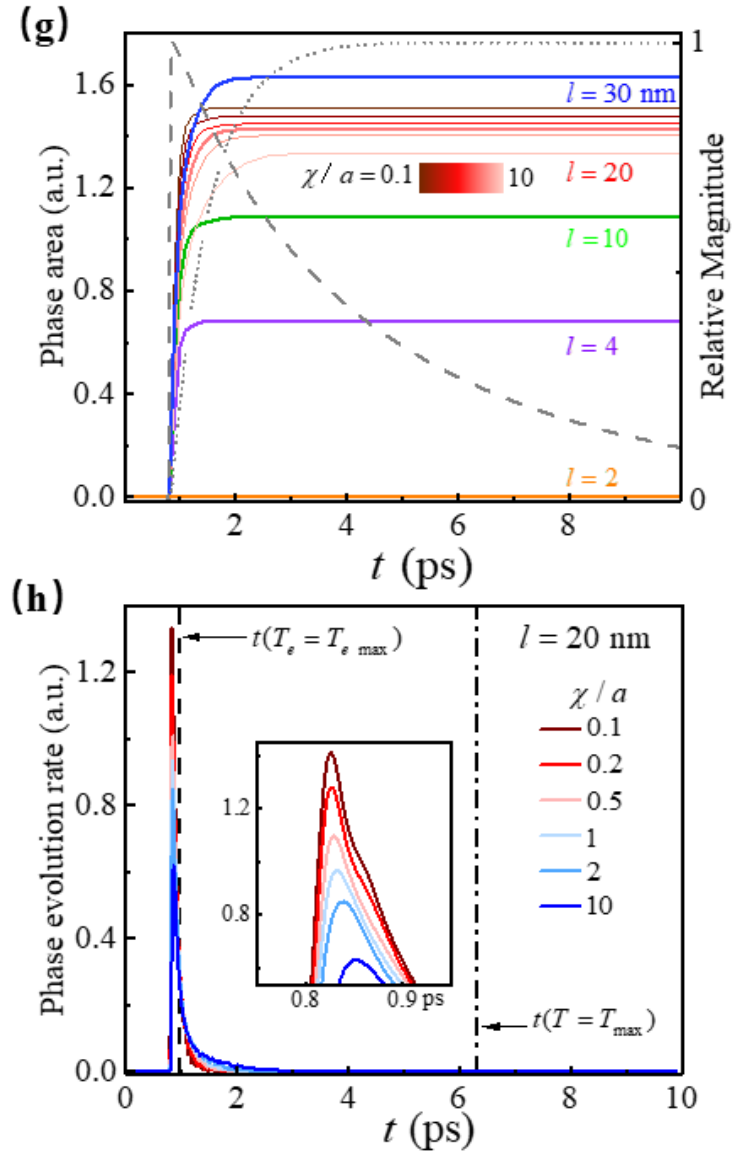

Figure 4 (a)-(c) Contour map of the β field and (d)-(f) the phase filed at different times. (g) Phase area evolution and (h) its time derivative with different particle size and critical distance.

### 3.3 Final state DPF

As previously described, the dynamic phase field (DPF) model rapidly evolves to a steady state where the phase interface arrests. This steady state is hereafter referred to as the final state. To characterize parameter-dependent final states, we analyze results from the last simulation time step $t = 10$ ps. Figure 5(a)-(e) shows the final state DPF of some typical parameters. Like some experimental results[23], [27]–[29], [35], a disc-shaped damage crater forms with approximately constant width. The crater depth exhibits distinct trends: it increases rapidly with equivalent particle size $l$ due to the reduced energy requirement for damaging unit solid volume with larger particles, while decreasing gradually with critical distance $\chi$ as a consequence of phase evolution kinetics.

Sweeping of 246 parameter points is shown in Figure 5(f), reveals the final state phase area $A_\infty$ follows specific relationships: it is a monotonic increasing convex function of $l$, and decreases approximately linearly with $\log_2(\chi)$. Particle size demonstrates a dominant influence: even over a 2-order-of-magnitude $\chi$ sweep from $10a$ to $0.1a$, $A_\infty$ increases only 17.4% for $l = 30$ nm. In contrast, $l$ increase from 2.5 nm to 5 nm at fixed $\chi = 0.1a$ produces a 54.3% $A_\infty$ enlargement. As expected, Consistently, the final phase area $A_\infty$ remains negligible for particle sizes $l \leq 2$ nm. At the specified laser energy density, damage initiation occurs exclusively when emitted particles exceed an equivalent critical size threshold, highlighting the size-dependent nature of damage formation under these conditions.

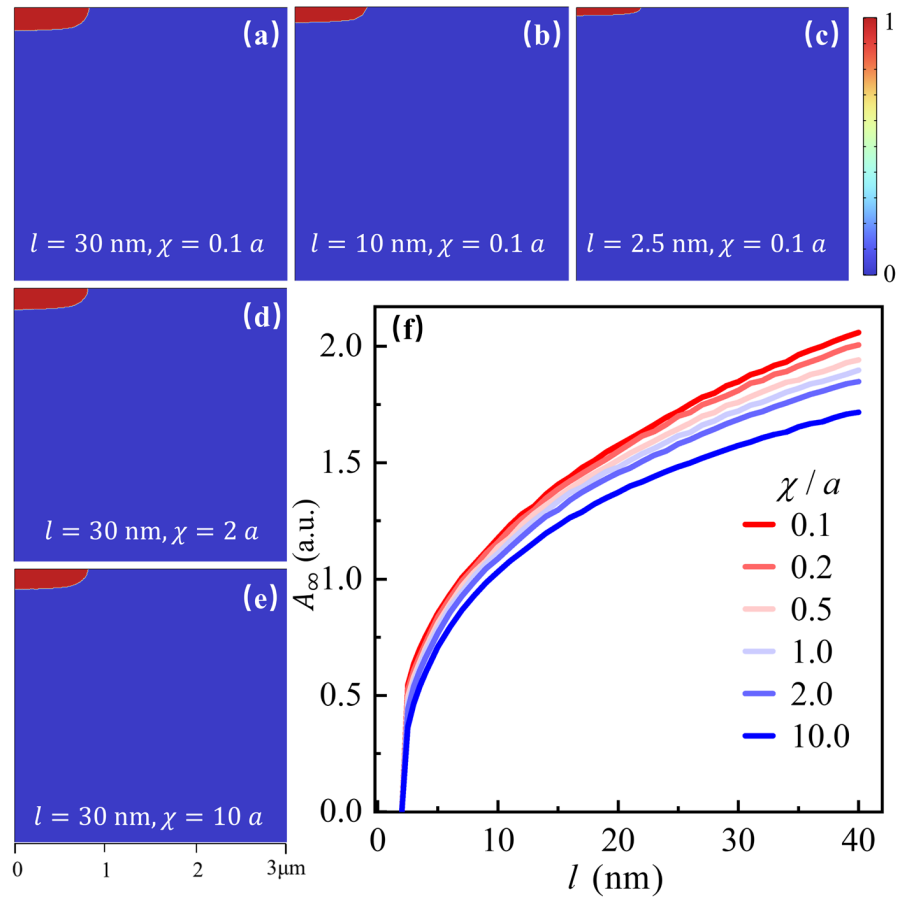

Figure 5 (a)-(c) Final state DPF for $\chi = 0.1a$ with different $l$. (a), (d), (e) Final state DPF for $l = 30$ nm with different $\chi$. (f) Final state DPF area $A_\infty$ as a function of $l$ and $\chi$.

## 4. Conclusion

We present a theoretical framework describing the solid damage evolution dynamics driven by particle emission during Coulomb explosion, elucidating the mechanism by which ultrashort laser pulses—characterized by low fluence yet high power density that insufficient to vaporize material—induce surface damage. Through four physically motivated approximations linking Allen-Cahn parameters to equivalent particle size $l$ and critical distance $\chi$, the model resolves the low-energy non-melting ablation mechanism and enables predictive damage evolution simulations using these intuitive parameters as inputs. Numerical validation confirms that DPF solutions evolve to a stable state on a time scale commensurate with electron-ion thermal relaxation, about 1 ps in our example. A quantitative study establishes scaling laws for damage morphology under Coulomb explosion conditions: the final damage phase area $A_\infty$ exhibits stronger dependence on $l$ than on $\chi$, and the damage phase growth is completely suppressed when $l \leq 2$ nm.

In future studies, the model could be extended by incorporating the phase field's coupling effects on electromagnetic fields, electron density, and temperature, thereby enabling its application to multi-pulse and long-pulse laser scenarios. Additionally, the damage phase field model could be generalized to high-energy-density conditions to investigate the competitive interplay between Coulomb explosion and thermal effects, elucidating the dominant mechanisms. Furthermore, experimental validation could be conducted to establish the functional dependence of Allen-Cahn equation parameters on observable quantities, enhancing the model's predictive capability and physical relevance.

**Funding.** No funding supports this work.

**Acknowledgment.** The research uses the computing resources provided by College of Physics and Optoelectronic, Shenzhen University.

**Disclosures**. The authors declare no conflicts of interest.

## List of symbols

| Spatial vector fields | | |
|---|---|---|
| | meaning | unit |
| $E$ | Electric field | V/m |

| $H$ | Magnetic field | B |
|---|---|---|
| $J_e$ | Polarization current density | A/m$^2$ |
| **Spatial scalar fields** | | |
| | meaning | unit |
| $x, y$ | Two-dimensional space coordinates | m |
| $n_e$ | Excited electron density | m$^{-3}$ |
| $T_e$ | Electron temperature | K |
| $T$ | Ion temperature | K |
| $I$ | Laser intensity | W/m$^2$ |
| $\epsilon_1$ | dielectric permittivity (real parts) | 1 |
| $\epsilon_2$ | dielectric permittivity (imaginary parts) | 1 |
| $\phi$ | Damage phase | 1 |
| $\alpha$ | Barrier parameter | J/m$^3$ |
| $\beta$ | Free energy release parameter | J/m$^3$ |
| $M$ | Phase interface mobility | m$^3$/J/s |
| $\kappa$ | Interface thickness control coefficient | J/m |
| $\epsilon_e$ | Electron internal energy density | J/m$^3$ |
| $p_e$ | Electron pressure | J/m$^3$ (Pa) |
| **Variable scalar** | | |
| | meaning | Used value |
| $\lambda$ | Incident laser wavelength | 600 nm |
| $\omega$ | Incident laser angle frequency | $3.14 \times 10^{15}$ rad/s |
| $w_0$ | Gaussian beam waist width | 2 μm |
| $t_p$ | Gaussian pulse center time | 800 fs |
| $\epsilon_r$ | Relative dielectric permittivity | 2.105 |
| $n_0$ | Initial refractive index | 1.45 |
| $\tau_e$ | Electron collision relaxation time | 0.5 fs |
| $\tau_{rec}$ | Electron recombination relaxation time | 6 ps |
| $\tau_{ei}$ | Electron-phonon interaction relaxation time | 1 ps |
| $n_a$ | Saturation electron density | $2 \times 10^{-28}$ m$^{-3}$ |
| $\chi_3$ | Kerr coefficient of third order | $2 \times 10^{-22}$ m$^2$V$^{-2}$ |
| $D_e$ | Electron self-diffusion coefficient | $5400/(T/K)$ cm$^{-2}$s$^{-1}$ |
| $\mu_e$ | Electron mobility | $3 \times 10^{-5}$ m$^2$/V/s |
| $C_i$ | Specific heat capacity of the solid | 703 J/kg/K |
| $\rho$ | Density of the solid | 2203 kg/m$^3$ |
| $a$ | Lattice constant | 0.5 nm |
| $l_i$ | Mesh size at the phase interface | 5 nm |
| **Constant** | | |
| | meaning | Value |
| $\epsilon_0$ | Vacuum dielectric constant | $8.854 \times 10^{-12}$ F/m |
| $\mu_0$ | Vacuum permeability | $4\pi \times 10^{-7}$ N/A$^2$ |
| $c$ | Speed of Light | $2.997 \times 10^8$ m/s |
| $e$ | Elementary charge | $1.602 \times 10^{-19}$ C |
| $m_e$ | Mass of electron | $9.109 \times 10^{-31}$ kg |

| $k_B$ | Boltzmann constant | $1.38 \times 10^{-23}$ J/K |

**Defined nouns and abbreviations**

| | meaning |
|---|---|
| DPF | Damage phase field |
| I-H | Ionization-heating |
| *critical distance* | A distance between an emitted particle and the solid surface. When the particle is emitted up to this distance, the electron repulsion pressure between them jumps from $p_e$ to 0 |
| *final state* | During Coulomb explosion, the DPF quickly evolves into such a steady state, the phase interface stops moving. |

Reference

[1] K. Xu, L. Huang, and S. Xu, "Line-shaped laser lithography for efficient fabrication of large-area subwavelength nanogratings," *Optica*, vol. 10, no. 1, p. 97, 2023, doi: 10.1364/optica.472730.

[2] N. Bonod, P. Brianceau, and J. Neauport, "Full-silica metamaterial wave plate for high-intensity UV lasers," *Optica*, vol. 8, no. 11, p. 1372, 2021, doi: 10.1364/optica.434662.

[3] F. Dausingera, "From basic understanding to technical applications," *Physics (College. Park. Md).*, vol. 5147, pp. 106–115, 2003.

[4] A. Žemaitis, M. Gaidys, P. Gečys, G. Račiukaitis, and M. Gedvilas, "Rapid high-quality 3D micro-machining by optimised efficient ultrashort laser ablation," *Opt. Lasers Eng.*, vol. 114, no. October 2018, pp. 83–89, 2019, doi: 10.1016/j.optlaseng.2018.11.001.

[5] J. Serbin, T. Bauer, C. Fallnich, A. Kasenbacher, and W. H. Arnold, "Femtosecond lasers as novel tool in dental surgery," *Appl. Surf. Sci.*, vol. 197–198, pp. 737–740, 2002, doi: 10.1016/S0169-4332(02)00402-6.

[6] V. Lekarev *et al.*, "Mechanism of lithotripsy by superpulse thulium fiber laser and its clinical efficiency," *Applied Sciences (Switzerland)*, vol. 10, no. 21. pp. 1–11, 2020. doi: 10.3390/app10217480.

[7] S. Egawa *et al.*, "Observation of mammalian living cells with femtosecond single pulse illumination generated by a soft X-ray free electron laser," *Optica*, vol. 11, no. 6, pp. 736–743, Jun. 2024, doi: 10.1364/OPTICA.515726.

[8] M. M. Günther *et al.*, "Forward-looking insights in laser-generated ultra-intense γ-ray and neutron sources for nuclear application and science," *Nat. Commun.*, vol. 13, no. 1, pp. 1–13, 2022, doi: 10.1038/s41467-021-27694-7.

[9] B. J. Demaske, V. V. Zhakhovsky, N. A. Inogamov, C. T. White, and I. I. Oleynik, "MD simulations of laser-induced ultrashort shock waves in nickel," *AIP Conf. Proc.*, vol.

1426, no. March, pp. 1163–1166, 2012, doi: 10.1063/1.3686486.

[10] R. F. W. Herrmann, J. Gerlach, and E. E. B. Campbell, "Ultrashort pulse laser ablation of silicon: An MD simulation study," *Appl. Phys. A Mater. Sci. Process.*, vol. 66, no. 1, pp. 35–42, 1998, doi: 10.1007/s003390050634.

[11] Y. Gan and J. K. Chen, "Numerical analysis of ultrashort pulse laser-induced thermomechanical response of germanium thin films," *Nanoscale Microscale Thermophys. Eng.*, vol. 16, no. 4, pp. 274–287, 2012, doi: 10.1080/15567265.2012.735350.

[12] M. Maigler *et al.*, "Atomistic modelling of ultrafast laser-induced melting in copper," vol. 12939, p. 61, 2024, doi: 10.1117/12.3012630.

[13] X. Liu, W. Zhou, C. Chen, L. Zhao, and Y. Zhang, "Study of ultrashort laser ablation of metals by molecular dynamics simulation and experimental method," *J. Mater. Process. Technol.*, vol. 203, no. 1–3, pp. 202–207, 2008, doi: 10.1016/j.jmatprotec.2007.09.084.

[14] F. J. Gürtler, M. Karg, K. H. Leitz, and M. Schmidt, "Simulation of laser beam melting of steel powders using the three-dimensional volume of fluid method," *Phys. Procedia*, vol. 41, pp. 881–886, 2013, doi: 10.1016/j.phpro.2013.03.162.

[15] C. Zenz *et al.*, "A compressible multiphase Mass-of-Fluid model for the simulation of laser-based manufacturing processes," *Comput. \& FLUIDS*, vol. 268, Jan. 2024, doi: 10.1016/j.compfluid.2023.106109.

[16] B. F. Akers and J. A. Reeger, "Numerical simulation of thermal blooming with laser-induced convection," *J. Electromagn. WAVES Appl.*, vol. 33, no. 1, pp. 96–106, 2019, doi: 10.1080/09205071.2018.1528183.

[17] J. Zhuang *et al.*, "Simulation on Plasma-Based Polarization Optics for Relativistic Laser Pulses," *ACTA Opt. Sin.*, vol. 45, no. 2, Jan. 2025, doi: 10.3788/AOS241466.

[18] C. Li, S. R. Vatsya, and S. K. Nikumb, "Effect of plasma on ultrashort pulse laser material processing," *J. Laser Appl.*, vol. 19, no. 1, pp. 26–31, 2007, doi: 10.2351/1.2402521.

[19] V. Morel, A. Bultel, I. Schneider, and C. Grisolia, "State-to-state modeling of ultrashort laser-induced plasmas," *Spectrochim. Acta - Part B At. Spectrosc.*, vol. 127, pp. 7–19, 2017, doi: 10.1016/j.sab.2016.11.002.

[20] M. E. Povarnitsyn, K. V. Khishchenko, and P. R. Levashov, "Phase transitions in femtosecond laser ablation," *Appl. Surf. Sci.*, vol. 255, no. 10, pp. 5120–5124, 2009, doi: 10.1016/j.apsusc.2008.07.199.

[21] M. E. Povarnitsyn, P. R. Levashov, and K. V. Khishchenko, "Implementation of kinetics of phase transitions into hydrocode for simulation of laser ablation," *High-Power Laser Ablation VII*, vol. 7005, p. 700508, 2008, doi: 10.1117/12.782581.

[22] M. Saghebfar, M. K. Tehrani, S. M. R. Darbani, and A. E. Majd, "Femtosecond pulse laser ablation of chromium: experimental results and two-temperature model simulations," *Appl. Phys. A Mater. Sci. Process.*, vol. 123, no. 1, pp. 1–9, 2017, doi: 10.1007/s00339-016-0660-0.

[23]   A. V. Bulgakov, J. Sládek, J. Hrabovský, I. Mirza, W. Marine, and N. M. Bulgakova, "Dual-wavelength femtosecond laser-induced single-shot damage and ablation of silicon," *Appl. Surf. Sci.*, vol. 643, no. July 2023, 2024, doi: 10.1016/j.apsusc.2023.158626.

[24]   N. Stojanovic, "Laser ablation driven by femtosecond optical and XUV pulses," 2008.

[25]   C. Phipps, W. Bohn, T. Lippert, A. Sasoh, W. Schall, and J. Sinko, "A review of Laser Ablation Propulsion," *AIP Conf. Proc.*, vol. 1278, no. January 1998, pp. 710–722, 2010, doi: 10.1063/1.3507164.

[26]   T. Q. Jia *et al.*, "Theoretical and experimental study on femtosecond laser induced damage in CaF2 crystals," *Appl. Phys. A Mater. Sci. Process.*, vol. 81, no. 3, pp. 645–649, 2005, doi: 10.1007/s00339-004-2685-z.

[27]   D. Soto-Puebla, J. A. Parada-Peralta, and S. Alvarez-Garcia, "Visualizing femtosecond-laser processed graphene micropatterns through AFM phase and multi-parametric Raman mapping," *Appl. Surf. Sci.*, vol. 685, no. December 2024, p. 162051, 2025, doi: 10.1016/j.apsusc.2024.162051.

[28]   V. Temnov *et al.*, "Microscopic characterization of ablation craters produced by femtosecond laser pulses," in *High-Power Laser Ablation IV*, 2002, p. 1032. doi: 10.1117/12.482062.

[29]   Q. Zhang *et al.*, "Growth of laser damage in fused silica and CaF2 under 263 nm laser irradiation," *Appl. Phys. B Lasers Opt.*, vol. 130, no. 8, pp. 1–8, 2024, doi: 10.1007/s00340-024-08287-w.

[30]   V. A. Volodin *et al.*, "Single-shot selective femtosecond and picosecond infrared laser crystallization of an amorphous Ge/Si multilayer stack," *Opt. Laser Technol.*, vol. 161, no. January, 2023, doi: 10.1016/j.optlastec.2023.109161.

[31]   I. Mirza *et al.*, "Non-thermal regimes of laser annealing of semiconductor nanostructures: crystallization without melting," *Front. Nanotechnol.*, vol. 5, no. October, pp. 1–12, 2023, doi: 10.3389/fnano.2023.1271832.

[32]   K. Sokolowski-Tinten, J. Bialkowski, and D. Von Der Linde, "Ultrafast laser-induced order-disorder transitions in semiconductors," *Phys. Rev. B*, vol. 51, no. 20, pp. 14186–14198, 1995, doi: 10.1103/PhysRevB.51.14186.

[33]   H. Sopha *et al.*, "Laser-induced crystallization of anodic TiO2nanotube layers," *RSC Adv.*, vol. 10, no. 37, pp. 22137–22145, 2020, doi: 10.1039/d0ra02929g.

[34]   S. K. Sundaram and E. Mazur, "Inducing and probing non-thermal transitions in semiconductors using femtosecond laser pulses," *Nat. Mater.*, vol. 1, no. 4, pp. 217–224, 2002, doi: 10.1038/nmat767.

[35]   B. Chimier *et al.*, "Damage and ablation thresholds of fused-silica in femtosecond regime," *Phys. Rev. B - Condens. Matter Mater. Phys.*, vol. 84, no. 9, pp. 1–10, 2011, doi: 10.1103/PhysRevB.84.094104.

[36]   B. Stuart, M. Feit, S. Herman, A. Rubenchik, B. Shore, and M. Perry, "Nanosecond-to-femtosecond laser-induced breakdown in dielectrics," *Phys. Rev. B - Condens. Matter*

*Mater. Phys.*, vol. 53, no. 4, pp. 1749–1761, 1996, doi: 10.1103/PhysRevB.53.1749.

[37] F. Costache and J. Reif, "Femtosecond laser induced Coulomb explosion from calcium fluoride," *Thin Solid Films*, vol. 453–454, pp. 334–339, 2004, doi: 10.1016/j.tsf.2003.11.096.

[38] R. Annou and V. K. Tripathi, "Femtosecond laser pulse induced coulomb explosion," *34th EPS Conf. Plasma Phys. 2007, EPS 2007 - Europhys. Conf. Abstr.*, vol. 31, no. 3, pp. 1777–1780, 2007.

[39] C. Chenard-Lemire, L. J. Lewis, and M. Meunier, "Laser-induced Coulomb explosion in C and Si nanoclusters: The determining role of pulse duration," *Appl. Surf. Sci.*, vol. 258, no. 23, pp. 9404–9407, 2012, doi: 10.1016/j.apsusc.2011.11.022.

[40] S. Li *et al.*, "Possible evidence of Coulomb explosion in the femtosecond laser ablation of metal at low laser fluence," *Appl. Surf. Sci.*, vol. 355, pp. 681–685, 2015, doi: 10.1016/j.apsusc.2015.07.136.

[41] V. I. Mazhukin, M. M. Demin, A. V. Shapranov, and A. V. Mazhukin, "Role of electron pressure in the problem of femtosecond laser action on metals," *Appl. Surf. Sci.*, vol. 530, no. July, p. 147227, 2020, doi: 10.1016/j.apsusc.2020.147227.

[42] K. Park, M. Fernandino, and C. A. Dorao, "Thermal two-phase flow with a phase-field method," *Int. J. Multiph. Flow*, vol. 100, pp. 77–85, 2018, doi: 10.1016/j.ijmultiphaseflow.2017.12.005.

[43] S. Guo, W. Liu, Q. Yang, X. Qi, and D. Yun, "Phase Field Simulation of Viscous Sintering," *Jinshu Xuebao/Acta Metall. Sin.*, vol. 60, no. 12, pp. 1691–1700, 2024, doi: 10.11900/0412.1961.2022.00445.

[44] C. sheng Zhu, Y. Lei, P. Lei, Z. hao Gao, and B. rui Zhao, "Phase field simulation of single bubble behavior under magnetic field," *Chinese J. Phys.*, vol. 89, no. January 2023, pp. 820–833, 2024, doi: 10.1016/j.cjph.2023.10.029.

[45] M. Plapp, "Phase-field simulations of crystal growth," *AIP Conf. Proc.*, vol. 1270, no. July 2010, pp. 247–254, 2010, doi: 10.1063/1.3476229.

[46] V. I. Mazhukin, A. V Shapranov, A. V Mazhukin, and O. N. Koroleva, "Mathematical Formulation of a Kinetic Version of Stefan Problem for Heterogeneous Melting / Crystallization of Metals," *Math. Montisnigri*, vol. XXXVI, pp. 58–77, 2016.

[47] R. Prieler, D. Li, and H. Emmerich, "Nucleation and successive microstructure evolution via phase-field and phase-field crystal method," *J. Cryst. Growth*, vol. 312, no. 8, pp. 1434–1436, 2010, doi: 10.1016/j.jcrysgro.2009.09.022.

[48] P. Ye *et al.*, "3D Lithiophilic CuZrAg Metallic Glass Based-Current Collector for High-Performance Lithium Metal Anode," *Small*, vol. 19, no. 52, pp. 1–13, 2023, doi: 10.1002/smll.202304373.

[49] J. Wang, S.-Q. Shi, L.-Q. Chen, Y. Li, and T.-Y. Zhang, "Phase-field simulations of ferroelectric/ferroelastic polarization switching," *Acta Mater.*, vol. 52, no. 3, pp. 749–764, 2004, doi: https://doi.org/10.1016/j.actamat.2003.10.011.

[50] T. Park, B. Ahmed, and G. Z. Voyiadjis, "A review of continuum damage and plasticity in concrete: Part I – Theoretical framework," *Int. J. Damage Mech.*, vol. 31, no. 6, pp. 901–954, 2022, doi: 10.1177/10567895211068174.

[51] E. Barchiesi and N. Hamila, "Maximum mechano-damage power release-based phase-field modeling of mass diffusion in damaging deformable solids," *Zeitschrift fur Angew. Math. und Phys.*, vol. 73, no. 1, pp. 1–21, 2022, doi: 10.1007/s00033-021-01668-7.

[52] G. Z. Voyiadjis and N. Mozaffari, "Nonlocal damage model using the phase field method: Theory and applications," *Int. J. Solids Struct.*, vol. 50, no. 20–21, pp. 3136–3151, 2013, doi: 10.1016/j.ijsolstr.2013.05.015.

[53] S. B. Biner and S. Y. Hu, "Simulation of damage evolution in discontinously reinforced metal matrix composites: A phase-field model," *Int. J. Fract.*, vol. 158, no. 2, pp. 99–105, 2009, doi: 10.1007/s10704-009-9351-6.

[54] A. Rudenko, J. P. Colombier, and T. E. Itina, "Nanopore-mediated ultrashort laser-induced formation and erasure of volume nanogratings in glass," *Phys. Chem. Chem. Phys.*, vol. 20, no. 8, pp. 5887–5899, 2018, doi: 10.1039/c7cp07603g.

[55] A. Rudenko, J. P. Colombier, and T. E. Itina, "From random inhomogeneities to periodic nanostructures induced in bulk silica by ultrashort laser," *Phys. Rev. B*, vol. 93, no. 7, pp. 1–13, 2016, doi: 10.1103/PhysRevB.93.075427.

[56] L. Capuano, D. de Zeeuw, and G. R. B. E. Römer, "Towards a numerical model of picosecond laser-material interaction in bulk sapphire," *J. Laser Micro Nanoeng.*, vol. 13, no. 3, pp. 166–177, 2018, doi: 10.2961/jlmn.2018.03.0005.

[57] A. Rudenko, A. Han, and J. V. Moloney, "Trade-Off between Second- and Third-Order Nonlinearities, Ultrafast Free Carrier Absorption and Material Damage in Silicon Nanoparticles," *Adv. Opt. Mater.*, vol. 11, no. 2, pp. 1–7, 2023, doi: 10.1002/adom.202201654.

[58] A. Rudenko, H. Ma, V. P. Veiko, J. P. Colombier, and T. E. Itina, "On the role of nanopore formation and evolution in multi-pulse laser nanostructuring of glasses," *Appl. Phys. A Mater. Sci. Process.*, vol. 124, no. 1, pp. 1–11, 2018, doi: 10.1007/s00339-017-1492-2.

[59] A. Rudenko, P. Rosenow, V. Hasson, and J. V. Moloney, "Plasma-free water droplet shattering by long-wave infrared ultrashort pulses for efficient fog clearing," *Optica*, vol. 7, no. 2, p. 115, 2020, doi: 10.1364/optica.382054.

[60] Y. Liu, H. Fan, H. Liu, and Q. Dai, "Ultrafast optical ablation of gold nanoparticles: An electron dynamics model for coulomb explosion," *Surfaces and Interfaces*, vol. 56, no. October 2024, p. 105547, 2025, doi: 10.1016/j.surfin.2024.105547.

[61] X. Li *et al.*, "Ultrafast Coulomb explosion imaging of molecules and molecular clusters," *Chinese Phys. B*, vol. 31, no. 10, 2022, doi: 10.1088/1674-1056/ac89df.

[62] P. Martin *et al.*, "Subpicosecond study of carrier trapping dynamics in wide-band-gap crystals," *Phys. Rev. B - Condens. Matter Mater. Phys.*, vol. 55, no. 9, pp. 5799–5810, 1997, doi: 10.1103/PhysRevB.55.5799.

[63] H. M. Van Driel, "Kinetics of high-density plasmas generated in Si by 1.06- and 0.53- m picosecond laser pulses," *Phys. Rev. B*, vol. 35, no. 15, pp. 8166–8176, 1987, doi: 10.1103/PhysRevB.35.8166.

[64] A. Rämer, O. Osmani, and B. Rethfeld, "Laser damage in silicon: Energy absorption, relaxation, and transport," *J. Appl. Phys.*, vol. 116, no. 5, p. 53508, 2014, doi: 10.1063/1.4891633.

[65] N. M. Bulgakova, R. Stoian, and A. Rosenfeld, "Laser-induced modification of transparent crystals and glasses," vol. 40, 2010, doi: 10.1070/QE2010V040N11ABEH014445.

[66] M. J. Grote, "Non-reflecting boundary conditions for electromagnetic scattering," *Int. J. Numer. Model. Electron. Networks, Devices Fields*, vol. 13, no. 5, pp. 397–416, 2000, doi: 10.1002/1099-1204(200009/10)13:5<397::AID-JNM374>3.0.CO;2-5.

[67] J. N. Ed, "Downloaded 02 / 25 / 25 to 120 . 229 . 205 . 223 . Redistribution subject to SIAM license or copyright ; see https://epubs.siam.org/terms-privacy SCATTERING OF MAXWELL ' S EQUATIONS WITH A LEONTOVICH BOUNDARY CONDITION," vol. 59, no. 4, pp. 1322–1334, 1999.

[68] F. Colombini, V. Petkov, and J. Rauch, "Incoming and disappearing solutions for Maxwell's equations," *Proc. Am. Math. Soc.*, vol. 139, no. 6, pp. 2163–2173, 2011, doi: 10.1090/s0002-9939-2011-10885-2.

[69] H. Zhu, L. Chu, H. Lv, Q. Ye, S. Juodkazis, and F. Chen, "Ultrafast Laser Manipulation of In-Lattice Plasmonic Nanoparticles," *Adv. Sci.*, vol. 2402840, pp. 1–10, 2024, doi: 10.1002/advs.202402840.

[70] F. Costache, M. Ratzke, D. Wolfframm, and J. Reif, "Femtosecond laser ionization mass spectrometric analysis of layers grown by pulsed laser deposition," *Appl. Surf. Sci.*, vol. 247, no. 1–4, pp. 249–255, 2005, doi: 10.1016/j.apsusc.2005.01.064.

[71] V. F. Kovalev and V. Y. Bychenkov, "Relativistic Coulomb Explosion of a Spherical Microtarget," *Bull. Lebedev Phys. Inst.*, vol. 50, no. October, pp. S762–S771, 2023, doi: 10.3103/S1068335623190089.

[72] M. Plapp, "Three-dimensional phase-field simulations of directional solidification," *J. Cryst. Growth*, vol. 303, no. 1 SPEC. ISS., pp. 49–57, 2007, doi: 10.1016/j.jcrysgro.2006.12.064.

[73] T. Takaki, "Phase-field modeling and simulations of dendrite growth," *ISIJ Int.*, vol. 54, no. 2, pp. 437–444, 2014, doi: 10.2355/isijinternational.54.437.

[74] V. I. Mazhukin, "Kinetics and Dynamics of Phase Transformations in Metals Under Action of Ultra-Short High-Power Laser Pulses," *Laser Pulses - Theory, Technol. Appl.*, 2012, doi: 10.5772/50731.

[75] V. I. Mazhukin, A. V. Shapranov, and A. V. Mazhukin, "The structure of the electrical double layer at the metal-vacuum interface," *Math. Montisnigri*, vol. 44, no. April, pp. 110–121, 2019, doi: 10.20948/mathmon-2019-44-9.

[76] J. Yang and J. Kim, "On a two-phase incompressible diffuse interface fluid model with curvature-dependent mobility," *J. Comput. Phys.*, vol. 525, p. 113764, 2025, doi: https://doi.org/10.1016/j.jcp.2025.113764.

[77] G. WANG, D. ZENG, and Z. LIU, "Phase field calculation of interface mobility in a ternary alloy," *Trans. Nonferrous Met. Soc. China*, vol. 22, no. 7, pp. 1711–1716, 2012, doi: https://doi.org/10.1016/S1003-6326(11)61377-0.

[78] V. Hatty, H. Kahn, and A. H. Heuer, "Fracture Toughness, Fracture Strength, and Stress Corrosion Cracking of Silicon Dioxide Thin Films," *J. Microelectromechanical Syst.*, vol. 17, no. 4, pp. 943–947, 2008, doi: 10.1109/JMEMS.2008.927069.

[79] H. K. Malik and L. Devi, "Self-defocusing of super-Gaussian laser beam in tunnel ionized plasmas," *Optik (Stuttg).*, vol. 222, no. May, p. 165357, 2020, doi: 10.1016/j.ijleo.2020.165357.